\documentclass[superscriptaddress,amsmath,amssymb,aps,prl,twocolumn,floatfix]{revtex4}

\usepackage{cancel}
\usepackage[normalem]{ulem}
\usepackage{graphicx}
\usepackage{dcolumn}
\usepackage{bm}
\usepackage[colorlinks,citecolor=magenta]{hyperref}
\usepackage{lipsum}
\usepackage[utf8]{inputenc}
\usepackage{siunitx}

\begin{document}

\title{Three-photon excitation of InGaN quantum dots}

\author{Viviana Villafañe}
\thanks{Contributed equally to this work.}
\author{Bianca Scaparra}
\thanks{Contributed equally to this work.}
\author{Manuel Rieger}
\author{Stefan Appel}
\affiliation{Walter Schottky Institut and Physik Department, Technische Universit{\" a}t M{\" u}nchen, Am Coulombwall 4, 85748 Garching, Germany}
\author{Rahul Trivedi}
\affiliation{Max-Planck-Institute for Quantum Optics, Hans-Kopfermann-Str. 1, 85748 Garching, Germany}
\author{Tongtong Zhu}
\author{John Jarman}
\author{Rachel A. Oliver}
\affiliation{Department of Materials Science, University of Cambridge, Cambridge, United Kingdom of Great Britain and Northern Ireland}
\author{Robert A. Taylor}
\affiliation{Clarendon Laboratory, Department of Physics, University of Oxford, Oxford OX1 3PU, United Kingdom}
\author{Jonathan J. Finley}
\author{Kai M\"uller}
\email{kai.mueller@wsi.tum.de}
\affiliation{Walter Schottky Institut and Physik Department, Technische Universit{\" a}t M{\" u}nchen, Am Coulombwall 4, 85748 Garching, Germany}

\begin{abstract}

We demonstrate that semiconductor quantum dots can be excited efficiently in a resonant three-photon process, whilst resonant two-photon excitation is highly suppressed. Time-dependent Floquet theory is used to quantify the strength of the multi-photon processes and model the experimental results. The efficiency of these transitions can be drawn directly from parity considerations in the electron and hole wavefunctions in semiconductor quantum dots.
Finally, we exploit this technique to probe intrinsic properties of InGaN quantum dots. In contrast to non-resonant excitation, slow relaxation of charge carriers is avoided which allows us to measure directly the radiative lifetime of the lowest energy exciton states. Since the emission energy is detuned far from the resonant driving laser field, polarization filtering is not required and emission with a greater degree of linear polarization is observed compared to non-resonant excitation.
\end{abstract}

\maketitle

Optically-active few-level quantum systems play a pivotal role for probing fundamentals of light-matter interactions. In the solid-state, color centers and semiconductor quantum dots are prominent systems for these tasks due to their strong coupling to light. Coherent control of electrons in quantum dots via optically-tailored pulsed lasers is a widely used technique to induce a precise quantum evolution of the electrons under strong-field interaction\cite{dory2016complete,sbresny2021stimulated,press2008complete,cerfontaine2020closed,stievater2001rabi,zrenner2002coherent,ramsay2010review}.
These schemes demonstrate good agreement between theoretical predictions and experimental measurements, showing that quantum dots are good candidates to test quantum mechanical results. Thereby, we explore multi-photon absorption selection rules in indium gallium nitride (InGaN) semiconductor quantum dots. 
In general, multi-photon absorption selection rules
between two quantum levels depend on the symmetry of the
states involved in the excitation process and the light polarization. However, if the excitation scheme involves N-photons of the same energy and polarization, the dipole approximation predicts that either all even or odd resonances are enhanced, based on the parity of the ground and excited states. In the case for semiconductor quantum dots, odd-photon resonances are enhanced whilst resonant even-photon excitations are suppressed\cite{magnuson2010electronic,grynberg1977doppler,grynberg1979three,bonin1984two,PhysRev.138.B979}(see Supplementary Material (SM) for details, which contains Refs. \cite{boyd2020nonlinear,Barra,gontier1971multiphoton, lambropoulos1976topics,winkelnkemper2006interrelation,steck2007quantum,shirley1965solution,bastard1981superlattice,PhysRevLett.14.60,bassani1977choice,parthey2011improved,saleh1998entangled,moscatelli1986simple,saito2002electronic,jarjour2005two}).

Even though this selection rule arises from simple parity considerations on the involved states, its experimental confirmation on semiconductor quantum dots has remained elusive for decades. The key experimental challenges include (i) the need of a pristine semiconductor two-level system composed of a ground and a excited state without any intermediate real states; (ii) the prerequisite for high peak power intensities of the laser field to perform multi-photon absorption experiments; and (iii) the long wavelengths used in a multi-photon experiment as compared to the emission wavelength of the two-level system.

Semiconductor quantum dots (QDs) formed from group-III nitrides are a promising platform to prove the presented selection rule. Indium gallium nitride (InGaN) QDs are  particularly  attractive since  their emission energy lies in the blue and green spectral region, a range well-matched to commercially available ultrafast single photon detectors and easily reachable in resonant multi-photon absorption experiments \cite{sultana2015tunable,lee2017cubic}. InGaN QDs have also well developed epitaxial growth methods that permit the fabrication of pristine quantum systems that possess high brightness, fast radiative decay times and on-demand emission of single photons at elevated temperatures (200K)  \cite{C7NR03391E,patra2019exploring,zhang2016charge,kako2006gallium}. 
\begin{figure*}
\includegraphics[scale=0.55, trim=0 10 200 0, clip]{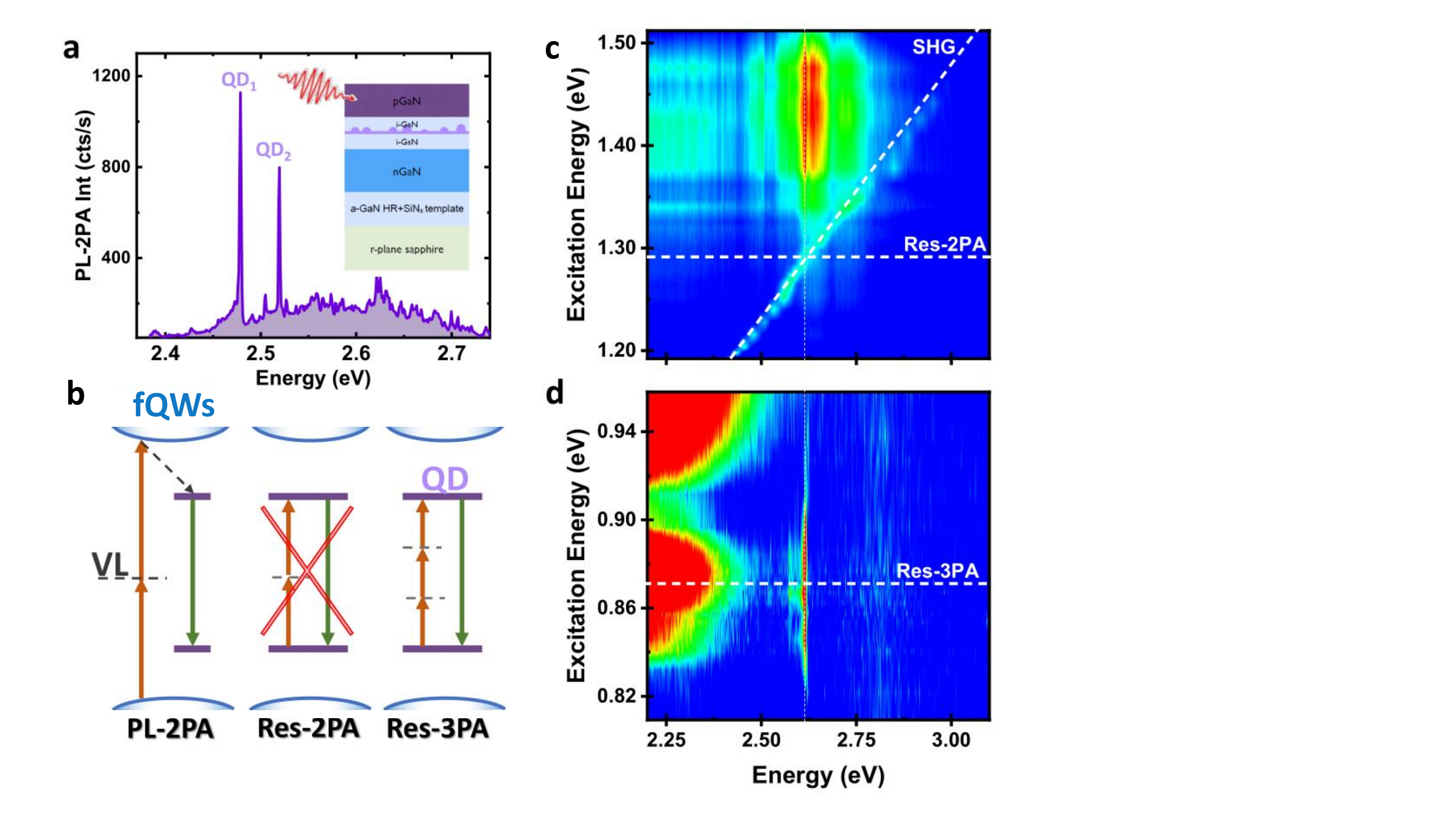}
\caption{\label{f1-2}
(a) Photoluminescence spectra acquired using 1.512\,eV of excitation energy, corresponding to a 2PA into the GaN fQWs. The inset shows a schematic of the sample structure and materials.
(b) Scheme of the different experiments presented in this manuscript. Left panel: 2PA into the fQWs of the GaN semiconductor. Center panel (Right panel): 2PA (3PA) resonant to the QD levels inside the GaN gap. The resonant 3PA condition is highlighted with a dashed horizontal line.
(c) (False Color Image) Measured amplitude of the QD intensity as a function of the laser excitation energy. For clarity, the dashed vertical line highlights the QD emission ($\sim$2.61\,eV) and the transverse dashed line the SHG from the GaN. The 2PA resonant condition, at $\sim$1.3\,eV, is also marked by an horizontal line and calls attention to the disappearance of the QD spectra in the vicinity of those energies.
(d) (False Color Image) Measured amplitude of the QD intensity as a function of the laser excitation energy for energies close to the 3PA in the QD. The horizontal dashed line highlights the resonant three photon excitation.
}
\end{figure*}
In this letter, we present an experimental demonstration of this selection rule by exploring resonant multi-photon excitation of individual InGaN quantum dots. We model our results using the interaction Hamiltonian in the dipole approximation to describe the system evolution, the simplest Hamiltonian describing an optically driven quantum dot, and obtain good qualitative agreement. Moreover, our results reveal new information about the optical properties of the InGaN QDs when subject to resonant excitation. The resonant character of the excitation allows us to measure directly the radiative lifetime, which under non-resonant excitation is obscured by slow carrier relaxation processes. Intrinsically different from standard resonant excitation of excitons and biexcitons, in our stratagem the QD emission energy is far detuned from the resonant driving laser field and thus no polarization filtering is required. Finally, we observe a higher degree of linear polarization compared to traditional off-resonant excitation commonly used for group-III nitride quantum dots. 

The QDs under study are non-polar (11-20) $a$-plane InGaN QDs grown by metal-organic vapour phase epitaxy (MOVPE) embedded within a p-i-n doped gallium nitride (GaN) matrix. Details of the quantum dot growth process are given in Ref.\cite{doi:10.1063/1.4812345}. InGaN QDs are positioned in the centre of a 50 nm thick intrinsic GaN layer, which is clad by a 600 nm thick layer of n-doped GaN and a 200 nm thick layer of p-doped GaN, as shown on the inset of Fig.\ref{f1-2}(a). Nanopillar structures of radius $\sim$150\,nm are fabricated surrounding the QDs to allow for increased photon extraction efficiencies. 
An ultrafast laser setup is used to study the multi-photon excitation of the InGaN QDs. Single nanopillars are excited by $\sim$100\,fs tunable laser pulses within the 0.8-1.51\,eV range generated by a mode-locked Ti:sapphire laser that seeds an optical parametric oscillator having a repetition frequency of 80MHz. 
The excitation laser is focused onto the sample through a Cassegrain objective lens ($25 \times$, 0.3 N.A.), with emission collected via the same objective. We use a reflective objective to avoid chromatic aberrations between the excitation and detection wavelengths. The excitation laser is separated from the measured PL using a dichroic mirror. Fig.\ref{f1-2}(a) presents a typical microluminescence spectrum ($\mu$-PL) showing emission from two single InGaN QDs embedded in a nanopillar at 4K. The spectra possesses two sharp features identified as single QDs in the green-blue spectral region and is obtained using an excitation energy of 800\,nm (1.55\,eV) corresponding to a two-photon absorption (PL-2PA) into the continuum bands arising from disordered fragmented quantum wells (fQWs) in the InGaN structure\cite{jarjour2005two,collins2009two}. The single lines have characteristic asymmetric lineshapes, indicative of zero phonon transitions from individual InGaN QDs with coupling to a continuum of acoustic phonons. The QD emission appears on top of a low intensity background emitted by the fragmented InGaN quantum wells in the sample \cite{doi:10.1063/1.4812345}. 

The quantum level scheme of the experiments performed in this work is depicted schematically in Fig.\ref{f1-2}(b). It consists of an InGaN QD having a ground and excited orbital states within the GaN gap. As previously stated, typically QD $\mu$-PL is obtained by performing a PL-2PA into the continuum states of the fQWs, depicted on the left panel. Alternatively, the ground and excited state of the QD system can also be resonantly coupled by driving either a two- or three-photon absorption process using a laser pulse (solid orange arrows) tuned to be resonant on virtual levels (VL) inside the QD, illustrated in the middle and left panels of Fig.\ref{f1-2}(b). 

Fig.\ref{f1-2}(c) shows the emission intensity of another single dot as the excitation energy laser is tuned throughout the two-photon resonance with the fQWs continuum in the GaN matrix. When the laser excitation energy $E_{\rm {L}}$ is greater than approximately half of the fQWs interband transition energy $E_{\rm {S}}/2$, the single QD emission is clearly observed in the spectra at $E_{QD}\sim$2.61\,eV. As $E_{\rm {L}}$ is decreased, the QD emission progressively reduces until it vanishes when $E_{\rm {L}}=E_{\rm {QD}}/2$, corresponding to the resonant 2PA into the QD. The striking disappearance of the QD emission is further emphasized by the presence of the second harmonic generation (SHG) of the fundamental photons that can clearly be observed in Fig.\ref{f1-2}(c) \cite{doi:10.1063/1.1316776,https://doi.org/10.1002/sca.4950230304}.
Interestingly, the intensity of the single QD emission exhibits a sharp resonance when approaching the resonant three photon condition, depicted in Fig.\ref{f1-2}(d) and highlighted with a dashed horizontal white line. Note that in this case we observe not only the QD emission but also the presence of a red-shifted PL signal at $\sim$2.25\,eV arising from defects in the GaN matrix \cite{doi:10.1063/5.0041608,doi:10.1063/5.0045019}. In contrast, for the case of the Res-2PA excitation we measure the QD signal on top of an underlying background that arises from InGaN quantum wells \cite{doi:10.1063/1.4812345}. As a consequence, the QD spectra obtained with a resonant 3PA present a clearer and sharper signature and a better signal-to-noise ratio. This result already highlights one striking advantage of our excitation strategy; whilst resonant excitation usually requires cross-polarized filtering in the optical setup to distinguish between excitation laser and the generated single photons, the non-linear three-photon excitation presented in this letter allows simple spectral filtering using only a dichroic mirror in the experimental setup. 

\begin{figure*}
\includegraphics[scale=0.45, trim=0 130 0 0, clip]{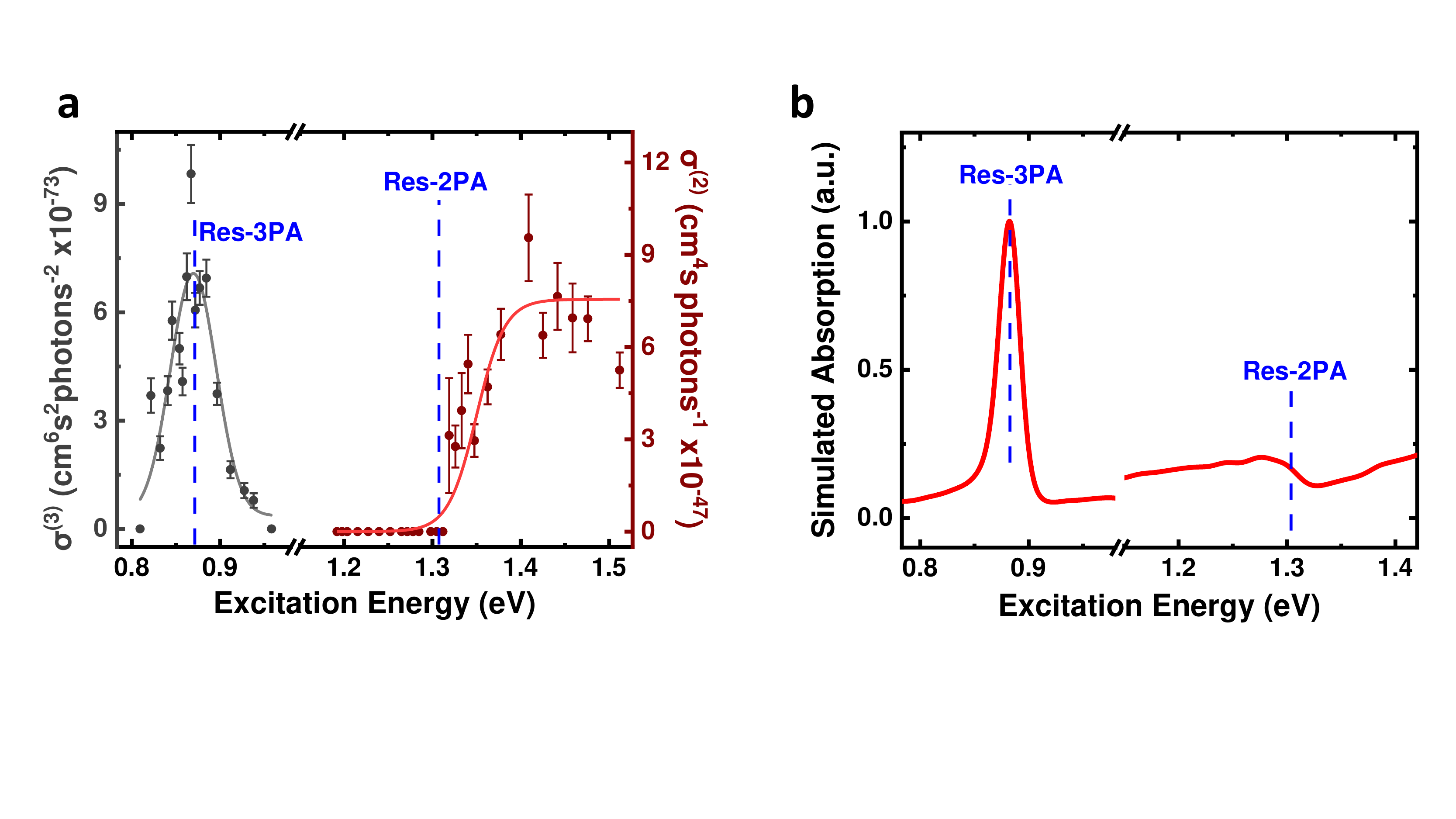}
\caption{\label{f3}
Experimental (a) and theoretical (b) absorption cross sections of a single InGaN QD as a function of the excitation energy. the resonant 2PA and 3PA conditions are highlighted by the blue vertical dashed lines. 
}
\end{figure*}

Fig.\ref{f3} summarizes the main results of our experiments, focusing on the detuning dependence of the amplitude of the InGaN QD PL. We first draw the reader\textsc{\char13}s attention to the experimental absorption cross-sections presented in Fig. \ref{f3}(a) and extracted from the data presented in Fig. \ref{f1-2}. Several features can be highlighted: (i) We observe a strong resonant enhancement of the QD signal close to to the 3PA condition. This resonance is energetically-sharp with a linewidth that corresponds to the spectral bandwidth of the excitation pulse (FWHM$\sim$20\,meV at 0.88\,eV). (ii) The QD PL intensity drops when tuning the laser energy to the Res-2PA energy, and (iii) the QD intensity rises again when performing a two-photon absorption into the continuum states of the QWs. 

It is worthwhile noticing that it is possible to measure a non-zero 2PA in typical semiconductor colloidal quantum dots, such as ZnSe and ZnSe/ZnS core-shell QDs. In these cases, the resonant 2PA absorption selection rule is determined by the superposition of molecular orbitals HOMO and LUMO (rather than discrete single-particle levels), the degree of electronic passivation in the QD surface and the localization of charge carriers in these core-shell quantum heterostructures\cite{lad2008two}. Conversely, CdSe colloidal dots present a resonant 2PA signal due to the presence of real intermediate states in between the excitation levels, that enhances the second order absorption coefficient \cite{makarov2014two}. Both of the aforementioned systems are not well-described by the electric dipole coupling of quantum states with opposite parities and without any real intermediate levels, and thus are out of the scope of the hypothesis made in this manuscript.

An interesting point to be made is that the experimental selection rule observed for InGaN QDs cannot be explained by arguments drawn from classical physics. As previously stated, the QDs presented in this work are grown along one of the non-polar planes of the Wurtzite group-III nitride. Consequently, this non-centrosymmetric system possesses a non-zero second-order susceptibility $\chi^{(2)}\sim1.3\times10^{-11}$\,m/V and a third order susceptibility $\chi^{(3)}\sim5.3\times10^{-19}$\,$\mathrm{m}^2/\mathrm{V}^2$ \cite{doi:10.1063/1.1316776}. 
Considering the power density used in the experiments of Fig.\ref{f1-2}, if our signal is being generated by SHG and THG processes in the fQWs, the ratio of electric dipole moments for the 2PA and 3PA excitation would be $P^{(2)}/P^{(3)}\sim350$. This would imply that the QD PL measured with a resonant 2PA excitation is much brighter than the one measured with a resonant 3PA, in strong contrast to our experimental findings.

To model our system theoretically and reproduce the measured two- an three-photon absorption intensities we use the interaction Hamiltonian in the electric dipole approximation considering the opposite parities of ground and excited states in the InGaN quantum dot as well as the existence of a second excited state $E_2$ for the electron in the QD (see SM) \cite{PhysRev.138.B979,saito2002electronic, jarjour2005two}: 
\begin{equation}
\label{eq2}
        {\rm i}\frac{{\rm d}}{{\rm d}t} \begin{pmatrix}a_g(t)\\a_1(t)\\a_2(t)\end{pmatrix} = \hat{\cal H}(t) \begin{pmatrix}a_g(t)\\a_1(t)\\a_2(t)\end{pmatrix},
    \end{equation} 
where
\begin{equation} 
\label{eq:3LS}
    \hat{\cal{H}}(t) = 
    \begin{pmatrix}
        E_g & 2b\sin(\omega t) & 0\\
        2b\sin(\omega t) & E_1 & 2b^\prime \sin(\omega t)\\
        0 & 2b^\prime \sin(\omega t) & E_2
    \end{pmatrix}.
\end{equation}

In Eq.\ref{eq2}, $\omega$ denotes the laser frequency and  $\omega_0 = (E_1-E_g)/\hbar$ the frequency difference between the ground and first excited state. We use $(E_2-E_1) \simeq \SI{70}{meV}$, a typical value found in InGaN QDs. $b$ and $b^\prime$ represent the strength of the dipole coupling between levels and  $a_g(t)$, $a_1(t)$, $a_2(t)$ are the probability amplitudes of occupation of the ground and excited states in the QD.
We calculate the quasienergies of the Hamiltonian $\hat{\cal H}$ for varying laser energy $\omega$ and constant ratios $b/\omega_0$ and $b/b^\prime$ using the Floquet solver of the Python package QuTiP \cite{JOHANSSON20121760,JOHANSSON20131234}. (see SM).

To estimate the electric field amplitude $b$, we calculate the transition dipole moment \cite{DipoleMoment-formula} for a typical measured radiative lifetime of \SI{250}{ps} 
    \begin{equation}
        |M_{12}| = \sqrt{\frac{3\epsilon_0 c^3 \hbar}{\tau \cdot 8 \pi^2 \omega_0^3 n_{\text{GaN}}}} \approx \SI{0.5}{\elementarycharge \nano\meter},
    \end{equation}
where $\hbar$ is Planck's constant, $\epsilon_0$ is the vacuum permittivity, $c$ is the speed of light and $n_{\text{GaN}}$ is the gallium nitride refractive index. We find that $|M_{12}|\sim\SI{0.5}{\elementarycharge \nano\meter}$, which is in the same order of magnitude as literature values for similar InGaN/GaN quantum dots \cite{DipoleMoments_InGaN}.
Considering that we use a pulsed laser, we have an average electric field amplitude, $E_0$, of $\sqrt{2 I_{\rm avg} / (c \epsilon_0 \cdot\SI{80}{MHz} \cdot \SI{100}{fs})} = \SI{2.2}{MV/cm}$ during the pulse.
The constant $b$ is half the Rabi frequency, which we calculate from the measured power density at three photon absorption ($\SI{283}{kW/cm^2}$) as $b = |M_{12}| \cdot E_0 / (2\hbar) = \SI{84}{THz}$. We approximate the ratio $b^\prime/b \simeq 0.44$ (see SM).

Panel (b) in Fig.~\ref{f3} presents the simulated absorption coefficient. Essentially, the simulation presents two peaks positioned at the Res-2PA and Res-3PA condition. The measured resonance is a convolution of the spectral width of the ultrashort pulses and the simulated absorption peaks. The model predicts an intensity $\sim$12 times smaller for the Res-2PA as compared to the Res-3PA condition. The simulation does not describe the rise in intensity resulting from the PL-2PA absorption into the fQWs continuum since it only models confined states in the QD system. 
Our calculations prove that the measured spectra are mostly determined by the amplitude probability of an electron to be promoted from the ground to the first excited state in the QD. It is remarkable that this simple and elegant Hamiltonian which describes many quantum systems is sufficient to describe the selection rule that we observe here experimentally.

\begin{figure}
\includegraphics[scale=0.375, trim=177 10 0 0, clip]{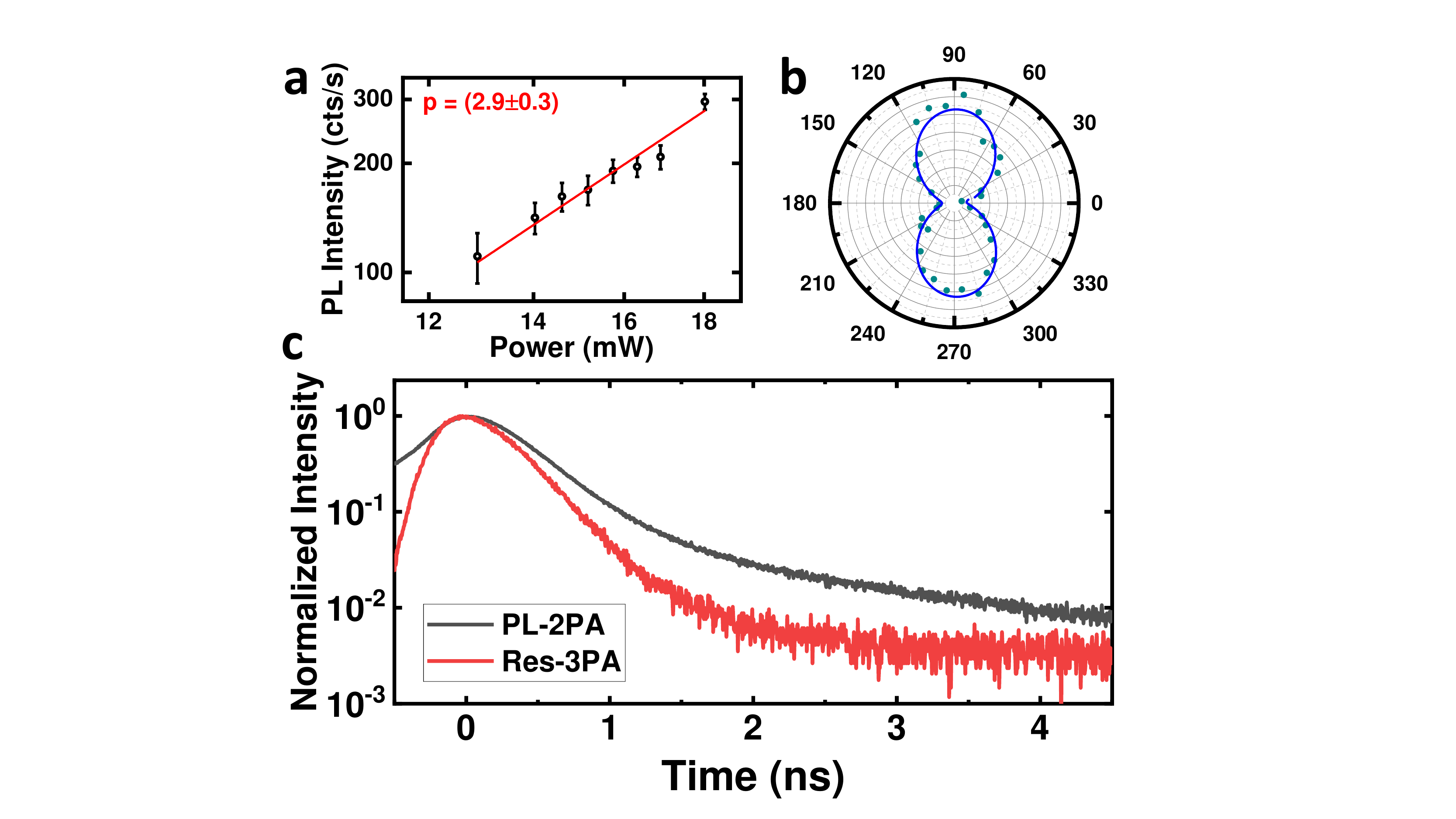}
\caption{\label{f4}
(a) Power dependence of the InGaN PL when performing a resonant 3PA excitation. We obtain a cubic characteristic exponent. (b) Polarization resolved measurement for the 3PA experiment. (c) Time-resolved PL for PL-2PA and the 3PA excitation. 
}
\end{figure}

To fully understand the nature of the spectra obtained with the Res-3PA condition, where $E_{\rm L}=E_{\rm QD}/3$, we explore the optical properties of the QD photon emission. Here we isolated the QD emission energy using tunable bandpass filters. Fig. \ref{f4} (a) shows the integrated PL intensity of the QD as a function of incident power in logarithmic scale. We observe a cubic characteristic power exponent equal to $(2.9\pm0.3)$, as expected for a 3PA process. We then performed polarization resolved measurements on the PL-2PA and Res-3PA excitation condition and compared the results. Sinusoidal fittings in accordance with Malus law show that the emission is linearly polarised in agreement with previous work, showing that not only the InGaN QDs present a linearly polarised emission, but also a deterministic polarisation axis along the $m$-direction of the nitride Wurzite system\cite{C7NR03391E}. Remarkably, the degree of linear polarization $\mathrm{DOLP} = (I_{\rm max}-I_{\rm min})/(I_{\rm max}+I_{\rm min})$ is improved from 74\% for the PL-2PA to 87\% when performing a resonant 3PA excitation on the same QD (Fig. \ref{f4}(b)). 

Finally, we investigate the time-resolved PL-intensity for the filtered QD signal for the PL-2PA and Res-3PA excitation conditions presented in Fig.\ref{f4}(c). We used a single-photon avalanche diode (SPAD) and a time-correlated single photon counting (TCSPC) module triggered by the laser to measure the time traces of the QD PL.
The radiative lifetime of the InGaN QD was obtained by fitting the temporal trace with a convolution between a Gaussian function and an exponential decay. The width of the Gaussian function was fitted in accordance with the instrument response function of the SPAD counter. The fitting gives an exponential component with a decay constant of $(440\pm20)$\,ps for PL-2PA experiment, decreasing to $(260\pm20)$\,ps in the 3PA resonant condition. The measured decay times are fast compared to the typical values obtained for $c$-plane InGaN QDs indicating that the QCSE in this sample is minimized. The observed decrease in the radiative decay time when performing a resonant excitation can be explained as follows: With a PL-2PA excitation we pump carriers into the fQW continuum, which then thermalize and can decay directly into the QD. Therefore, the extracted decay time reflects both the radiative exciton lifetime and slow carrier relaxation processes into the radiative state. In contrast, the Res-3PA case gives direct access to the true radiative lifetime since carrier relaxation does not take place.
This indicates that the resonant 3PA excitation scheme can be used to achieve faster repetition rates than the traditional PL-2PA excitation commonly used for InGaN QDs and, moreover, the associated jitter in the single photon emission events is minimized. Overall, the Res-3PA excitation improves the DOLP of the photonic emission and the radiative decay time up to the GHz regime.

In summary, we demonstrated experimentally that InGaN quantum dots can be efficiently excited by performing a resonant three-photon absorption, while resonant two-photon excitation is suppressed.
To this end, we presented spectroscopic measurements with a wide variety of laser excitation energies in a single InGaN QD and modelled the results using the Floquet Hamiltonian describing the system evolution. 
We expect our results to hold true for a wide variety of semiconductor quantum dots that posses the opposite parity in their ground and excited state wavefunctions.
We also showed that our excitation scheme involving the resonant 3PA of InGaN QDs enhances the degree of polarization and gives direct access to the intrinsic radiative lifetime of the QD. 
Our results shed new light on the fundamental quantum-mechanical selection rules describing semiconductor quantum dots and open up routes to implementing new resonant protocols for state preparation and quantum control.

\begin{acknowledgments}
We gratefully acknowledge financial support from the German Federal Ministry of Education and Research via the funding program Photonics Research Germany (contract number 13N14846), the BMBF via the project Q.Link.X (16KIS0874) and QR.X (16KISQ027) and the Deutsche Forschungsgemeinschaft (DFG, German Research Foundation) under Germany’s Excellence Strategy – EXC-2111 – 390814868. V. Villafañe gratefully acknowledges the MCQST Distinguished Fellowship program and the Alexander v. Humboldt foundation for financial support in the framework of their fellowship programs. R.A. Oliver and R.A. Taylor acknowledge the support of the EPSRC in the UK on grants EP/M012379/1 and EP/M011682/1. R. Trivedi acknowledges support from Max Planck Harvard research center for quantum optics (MPHQ).

\end{acknowledgments}

\bibliography{ref}

\end{document}